\documentclass[aps,nofootinbib,twocolumn,preprintnumbers,10pt]{revtex4}
\usepackage{amsmath,amssymb,graphicx,epsfig,breakurl,color,bm,subfigure,slashed,lipsum,appendix,multirow}
\usepackage{multirow}
\begin{document}

\title{ \Large
CP Violation of Baryon Decays with $N\pi$ Rescatterings }

\author{ Jian-Peng Wang\footnote{Email:wangjp20@lzu.edu.cn} and
Fu-Sheng Yu\footnote{Corresponding author, Email:yufsh@lzu.edu.cn}
\vspace{0.3cm}} 

\affiliation{
MOE Frontiers Science Center for Rare Isotopes, and School of Nuclear Science and Technology, Lanzhou University, Lanzhou 730000, China}

\begin{abstract}
There is a long-standing puzzle that the CP violation (CPV) in the baryon systems has never been well established in experiments, while the CPV of mesons have been observed by decades. In this paper, we propose that the CPV of baryon decays can be generated with the rescatterings of a nucleon and a pion into some final states, i.e. $N\pi\to N\pi$ or $N\pi\pi$.  Benefited by the fruitful data of $N\pi$ scatterings, we can model-independently analyse the strong phases of $b$-baryon decays using the partial wave amplitudes of $N\pi$ scatterings. Avoiding the most difficult problem of non-perturbative dynamics, it makes a great advantage to predict the CPV of baryon decays with a relatively reliable understanding of the decay dynamics. We study the processes of $\Lambda_b^0\to (p\pi^+\pi^-)h^-$ and $(p\pi^0)h^-$ with $h=\pi$ or $K$. It is found that the global CPV of the above processes in the invariant mass regions of $N\pi$ scatterings are at the order of several percent. More importantly, the local CPV in some regions of the Dalitz plots can reach the order of $10\%$, or be even larger. Considering the predicted results and the experimental data samples, we strong suggest to measure the CPV of $\Lambda_b^0\to (p\pi^+\pi^-)K^-$, which has a large possibility to achieve the first observation of CPV in the baryon system. 
\end{abstract}

\maketitle

%%%%%%%%%%%%%%%%%%%%%%%%%%%%%%%%%%
\section{Introduction}
The matter-antimatter asymmetry in the Universe requires the violation of the charge-parity (CP) symmetry in the microscopic particles and anti-particles \cite{Sakharov:1967dj}.
CP violation (CPV) has been observed in the $K$, $B$ and $D$ mesons sequentially during the past 60 years \cite{ParticleDataGroup:2022pth}. 
But it has never been well established in any baryon system, which is a long standing puzzle. 
%Besides, the CPV in the Standard Model of particle physics is not large enough to explain the matter-antimatter asymmetry in the Universe. 
In the fact that the visible matter of the universe is dominated by baryons, it is very important to study and search for CPV of baryon systems. 

In the Standard Model (SM) of particle physics, baryons do not mix with anti-baryons. CPV of baryons happens only in their decays, which is typically direct CPV.
It is well known that CPV always stems from the interference between tree and penguin diagrams in an individual process, such as $\Lambda_b^0\to p\pi^-$ or $pK^-$. 
However, it has not been observed for the CPV of the above processes even at the level of precision of $\mathcal{O}(1\%)$ \cite{LHCb:2018fly}.
In the multi-body decays, there exist a lot of resonances whose interference can also generate CPV. 
For example, in the three-body decay of $B^+\to\pi^+\pi^+\pi^-$, the interference between $f_0(500)$ and $\rho(770)$, or say the interference between the $S$-wave and $P$-wave amplitudes, contributes to the local CP violations as large as $(50-60)\%$ in some regions of the Dalitz plot \cite{LHCb:2019jta,LHCb:2022fpg}. 
Therefore, it deserves to investigate the multi-body decays of $b$-baryons.

A lot of baryon resonances can contribute to the multi-body decays of $b$-baryons. 
For instance, the excited states of nucleons $N^*$ with different spins and parities could provide fruitful effects of interference in the invariant masses of $p\pi$ or $p\pi\pi$. 
They might contribute to large strong phases, and thus possibly large regional CPV. 
However, there exists a problem meanwhile, that the number of excited $N^*$ states are too large. 
For example, it is recorded for more than 15 $N^*$ states in the mass region from the first excited state $N(1440)$ to  2 GeV in the Particle Data Group (PDG) \cite{ParticleDataGroup:2022pth}.
Their widths are usually larger than 100 MeV, making it very difficult to distinguish so many states within the invariant mass range of 600 MeV.
Besides, the masses and widths of these states are usually of large uncertainties in PDG. 
This makes a significant difficulty for both experimental partial-wave amplitude analysis and theoretical predictions. 

Actually, the excited $N^*$ states are usually determined in the experiments of $N\pi$ scatterings.  
Fortunately, scientists have analyzed the data of $N\pi$ scatterings and provided the partial-wave amplitudes. 
This can be easily obtained from the websites, such as the SAID program \cite{SAID}.
This helps us model-independently calculate the strong phases of $\Lambda_b$ decays, regardless of the detailed information of the resonant states in the invariant masses of $p\pi$ or $p\pi\pi$. 
It naturally solves the problem of too many resonances with large widths and large uncertainties from the excited $N^*$ states. 
The strong phases are usually of the most difficulty and of the largest uncertainty in the predictions of CPV in theory. 
But using the data of scatterings can significantly improve the theoretical precision on the predictions of CPV.
For example, the $\pi\pi\leftrightarrow K\bar K$ rescattering data leads to a very accurate description of CPV in $B^\pm\to K^\pm\pi^+\pi^-$ and $K^\pm K^+K^-$ \cite{AlvarengaNogueira:2015wpj,Cheng:2020ipp,Garrote:2022uub}, and the enhancement of CPV of $D^0\to K^+K^-$ and $\pi^+\pi^-$ \cite{Bediaga:2022sxw}.
Analogously, the partial-wave amplitudes obtained in the data of $N\pi\to p\pi$ or $p\pi\pi$ scatterings would be helpful to achieve a model-independent determination of strong phases, thus improve the precision of predictions on CPV of $\Lambda_b$ decays. 

In this work, we propose that there might be large local CPV in the decays of $\Lambda_b^0\to (p\pi^+\pi^-)h^-$ and $(p\pi^0)h^-$ with $h=\pi$ or $K$, in the mass region of $p\pi^+\pi^-$ or $p\pi^0$ re-scattered from $N\pi$. 
The mechanism of CPV in these processes is depicted in Fig. \ref{fig:CPVmechanism}.
In the short-distance weak interaction, $\Lambda_b^0$ transited into $N\pi$ with an emitted $\pi^-$ or $K^-$. 
Then $N\pi$ scatters into $p\pi^+\pi^-$ or $p\pi^0$ via a long-distance strong interaction. 
It can be model-independently obtained for the partial-wave amplitudes of scatterings of $N\pi \to p\pi^+\pi^-$ and $p\pi^0$ from the data analysis by the SAID program \cite{SAID}.
This is the most important feature of this proposal. 
The interference effect between the partial-wave amplitudes are different for the tree and penguin operators, which results in the strong phase differences between the tree and penguin amplitudes of $\Lambda_b$ decays. 
The strong phases from $N\pi$ scatterings are usually large. 
So it is possible to generate large regional CPV. 

This is the first proposal for the CPV mechanism of baryon decays stemming from $N\pi$ scatterings. 
On one hand, the strong phases are data driven and can thus be model-independently obtained, so that the theoretical predictions are more reliable. 
It realizes the remarkable dynamical predictions of CPV in $\Lambda_b$ decays with long-distance contributions which are usually very difficult in theory \cite{Lu:2009cm,Hsiao:2014mua,Zhu:2018jet}.
On the other hand, we can do the angular analysis since the data of $N\pi$ scatterings provides very comprehensive partial-wave amplitudes. 
The interference between different partial-wave amplitudes can be revealed in the angular distributions. 
From a detailed numerical analysis, we predict both the global and regional CPV of $\Lambda_b^0\to (p\pi^+\pi^-)h^-$ and $(p\pi^0)h^-$ with $h=\pi$ or $K$, in the mass regions of scatterings of $N\pi\to p\pi^+\pi^-$ and $p\pi^0$. 
It is found that the global CPV of these processes are several percent, while the regional CPV could reach as large as $(20-30)\%$.
It could be even larger for the CPV of some moments of $\cos\theta$ in the angular distributions.
Some suggestions are then given for the experimental measurements, which are of large possibility for the first observation of baryon CPV.

%%%%%%%%%%%%%%%%%%%%%%%%%%%%%%%%%%
\section{CPV mechanism with $N\pi$ scatterings}

\begin{figure}[!t]
\centering
%    \subfigure[]{    {\label{fig:ppipi_a}}
\includegraphics[scale=0.6]{"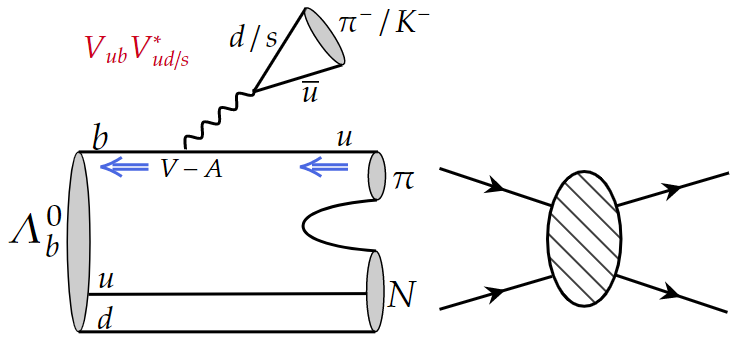"}\vspace{0.5cm}
%    \subfigure[]{    {\label{fig:pKpi_b}}
\includegraphics[scale=0.6]{"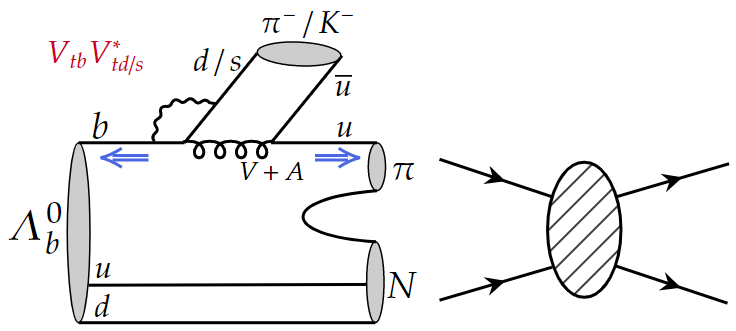"}
\caption{Illustration of the CPV mechanism driven by $N\pi$ re-scatterings. 
The processes undergo two steps, the short-distance weak interaction of $b$ decays firstly, and then long-distance strong interactions of final-state re-scatterings of $N\pi\to N\pi$ or $N\pi\pi$. 
The weak phases come from the tree and penguin diagrams, while the strong phases stem from the partial-wave amplitudes of $N\pi$ scatterings which can be model-independently obtained from the data of the SAID program. 
For example, the re-scatterings of $N\pi\to p\pi^0$ and $p\pi^+\pi^-$ contribute to the processes of $\Lambda_b^0\to (p\pi^+\pi^-)h^-$ and $(p\pi^0)h^-$ with $h=\pi$ or $K$, respectively. }\label{fig:CPVmechanism}
\end{figure}

The CPV mechanism of baryon decays via the $N\pi$ scatterings is depicted in Fig.\ref{fig:CPVmechanism}.
Take the decays of $\Lambda_b^0\to (p\pi^+\pi^-)h^-$ and $(p\pi^0)h^-$ as an example. 
The decay amplitudes of these processes undergo two steps,
\begin{equation}\label{eq:FSI}
\mathcal{A}=\mathcal{S}^{1/2}\mathcal{A}_0,
\end{equation}
given by the decay mechanism of final-state interaction of weak decays of heavy hadrons \cite{Suzuki:1999uc,Cheng:2004ru,Chua:2007cm}.
$\mathcal{A}_0$ is the short-distance weak decays of $\Lambda_b^0\to (N\pi)h^-$, while $\mathcal{S}^{1/2}$ represents the long-distance final-state re-scatterings of $N\pi\to p\pi^+\pi^-$ or $p\pi^0$. 
CPV comes from the interference between the diagrams of  \ref{fig:CPVmechanism} (a) and (b).
The weak phases are from the tree and penguin processes, while the strong phases stem from the $N\pi$ scatterings associated with the effective Wilson coefficients.

Firstly, $\Lambda_b$ decays into $(N\pi)h^-$ via a weak interaction of $b\to u\bar u d(s)$ at a very short time, i.e. the left-handed parts of Figs. \ref{fig:CPVmechanism} (a) and (b). 
This is a short-distance dynamics due to the large mass scale of $b$ quark and the weak interaction. 
The $N\pi$ system can have all possible quantum numbers of spin and parity. 
The decay amplitudes of $\Lambda_b\to (N\pi)h^-$ with short-distance dynamics are expressed as
\begin{align}\label{eq:A0}
\mathcal{A}_0=&\hspace{0.3cm}\bar u_{N\pi,1/2^+}(A+B\gamma_5)u_{\Lambda_b}
\nonumber\\
&+\bar u_{N\pi,1/2^-}(\tilde A+\tilde B\gamma_5)u_{\Lambda_b}
\nonumber\\
&+q_\mu\bar u_{N\pi,3/2^+}^\mu(C + D \gamma_5)u_{\Lambda_b}
\nonumber\\
&+q_\mu\bar u_{N\pi,3/2^-}^\mu(\tilde C + \tilde D \gamma_5)u_{\Lambda_b}
\nonumber\\
&+\cdots.
\end{align}
The spinors of final states correspond to the $N\pi$ system with different spins and parities, instead of that of the nucleon $N$. The $N\pi$ system is also a fermion which can be described by spinors. Such expression is more convenient to match with the $N\pi$ rescatterings in the following discussions. 
The spinors of spin-1/2 system is simple, while those of spin-3/2 are Rarita-Schwinger spinors with Lorentz indices.
$q_\mu$ is the four momentum of $h^-$.
The detailed expressions between the processes with the same spin but different parities are similar to each other, thus are distinguished only by tilde in the decay amplitudes of $A$, $B$ and $C$, $D$. 
It is noted that the above expressions are stringent, with only two terms with or without $\gamma_5$ in the decay amplitude for each process,  due to the simplicity of the emitted pseudoscalar meson $h=\pi$ or $K$.
The $N\pi$ system with higher spins are all given in a similar way, but not explicitly expressed here. 

A detailed dynamics of the decay amplitudes could be calculated under the factorization hypothesis.
It is well known that the factorization hypothesis works well for the external $W$-emission diagram, and this diagram is dominated compared to the other topological diagrams in the non-leptonic weak decays of heavy-flavor hadrons.
This can be seen in the decays of $B$-mesons \cite{Ali:1998eb}, $b$-baryons\cite{baryonPQCD}, $D$-mesons\cite{Li:2012cfa} and charmed baryons\cite{FSI-charm}.
In the factorization hypothesis, the decay amplitudes can be naively separated into the transition form factors of $\Lambda_b\to N\pi$ and the decay constants of emitted $h^-$ in both the tree and penguin diagrams, multiplied by the corresponding CKM matrix elements and effective Wilson coefficients.  
The full expressions of the $\Lambda_b\to N\pi$ form factors are long and not necessary to be explicitly given at the current stage. 
The form factors with the spin-parity states of $1/2^\pm$, $3/2^\pm$ and $5/2^\pm$ can be refereed to \cite{Mott:2011cx,Gutsche:2017wag,Meinel:2021mdj,Huang:2022lfr}. 
%
%\begin{align}&\langle N\pi|(\bar u b)_{\rm V-A}|\Lambda_b\rangle\nonumber\\&=\bar u_{N\pi}^{1/2^+}(f_1^{1/2^+}\gamma_\mu-g_1^{1/2^+}\gamma_\mu\gamma_5+\cdots)u_{\Lambda_b}\nonumber\\&+\bar u_{N\pi}^{1/2^-}(-f_1^{1/2^-}\gamma_\mu+g_1^{1/2^-}\gamma_\mu\gamma_5+\cdots)u_{\Lambda_b}\nonumber\\&+\bar u_{N\pi}^{3/2^+}(f_1^{3/2^+}\gamma_\mu-g_1^{3/2^+}\gamma_\mu\gamma_5+\cdots)u_{\Lambda_b}\nonumber\\&+\bar u_{N\pi}^{3/2^-}(-f_1^{3/2^-}\gamma_\mu+g_1^{3/2^-}\gamma_\mu\gamma_5+\cdots)u_{\Lambda_b}\nonumber\\&+\cdots,\end{align}
%

Secondly, the $N\pi$ system produced via short-distance weak interaction undergoes a long-distance re-scattering process, i.e. the right-handed parts of Figs. \ref{fig:CPVmechanism} (a) and (b). 
We only consider the re-scatterings of $N\pi\to p\pi^0$ and $p\pi^+\pi^-$ in this work to illustrate this novel CPV mechanism. 
There are fruitful experimental data of the scatterings of $N\pi\to N\pi$ and $N\pi\pi$. 
Scientists have provided the real and imaginary parts of the partial-wave amplitudes of these scatterings, as an open source in the website such as the SAID program \cite{SAID}.
The details of the partial-wave amplitudes can be found in the website, and shown with some examples in the Appendix \ref{sec:SAID}. 
This is the most important point of this work. 
The long-distance contributions are usually very difficult to be calculated due to the non-perturbative property. 
However, the use of the experimental data of scatterings could significantly reduce the theoretical uncertainties in the calculations of decay amplitudes. 
Besides, there are a lot of excited $N^*$ states whose masses and widths are of large uncertainties. 
The data of $N\pi$ scatterings would avoid the complicated and unclear information of these resonances.
The real and imaginary parts of the partial-wave amplitudes contribute to the strong phases which are necessary for CPV. 

In practice, $p\pi^+\pi^-$ is dominated by $\Delta^{++}\pi^-$ \cite{LHCb:2019jyj}. 
So we will discuss $\Delta^{++}\pi^-$ instead of $p\pi^+\pi^-$ for the convenience. It is found by the LHCb collaboration in \cite{LHCb:2019jyj} that the signal yields of $\Lambda^{0}_{b}\to \Delta^{++}\pi^{-}\pi^{-}$ and $\Delta^{++}\pi^{-}K^{-}$ are significantly large.  There are some other scattering processes, such as $N\pi\to N\rho\to p\pi\pi$, whose partial-wave amplitudes are also provided by the SAID program. However, we would like to leave the comprehensive analysis of $N\pi\to p\pi^+\pi^-$ in a detailed study in the near future. The purpose of this work is not to do a very comprehensive numerical predictions, but focus on the illustration of this new CPV mechanism via $N\pi$ scatterings. Therefore, it is good enough to use an approximation of the $\Delta^{++} \pi^-$ dominance in $p\pi^+\pi^-$ as a starting point.

Besides, the isospin of the $N\pi$ system can only be 1/2, since the initial state $\Lambda_b$ is isospin singlet, and the weak transition of $b\to u$ changes the isospin by $\Delta I=1/2$. 
Therefore, only the partial-wave amplitudes with isospin 1/2 contribute to the $N\pi$ scatterings in our case. 
We will use the same symbols as the SAID program to represent the partial-wave amplitudes, $L_{2I, 2J}$. 
$L$ is the angular momentum between $N$ and $\pi$, while $I$ and $J$ are the isospin and total spin of the system. 
For example, $D_{13}$ represents a partial-wave amplitude with $L=2$, $I=1/2$ and $J=3/2$. 
In fact, the spin-parity quantum states and the partial-wave amplitudes are one-to-one corresponding to each other, such as $1/2^+: P_{11}$, $1/2^-: S_{11}$, $3/2^+: P_{13}$, $3/2^-: D_{13}$, etc. 
Note that the partial-wave rescattering amplitudes depend on the invariant masses of $N\pi$.  
The details of the partial-wave amplitudes of $N\pi$ scatterings are shown in Appendix.\ref{sec:SAID} for the convenience. 
It can be found that the real and imaginary parts of the scattering amplitudes are of high precision, and thereby the strong phases can be precisely obtained.  
According to the final-state rescattering formula of Eq. (\ref{eq:FSI}), the total decay amplitudes are expressed as 
\begin{align}\label{eq:A0}
\mathcal{A}=&\hspace{0.3cm}\bar u_{N\pi,1/2^+}(A+B\gamma_5)u_{\Lambda_b} \;P_{11}
\nonumber\\
&+\bar u_{N\pi,1/2^-}(\tilde A+\tilde B\gamma_5)u_{\Lambda_b}\; S_{11}
\nonumber\\
&+q_\mu\bar u_{N\pi,3/2^+}^\mu(C + D \gamma_5)u_{\Lambda_b}\; P_{13}
\nonumber\\
&+q_\mu\bar u_{N\pi,3/2^-}^\mu(\tilde C + \tilde D \gamma_5)u_{\Lambda_b}\; D_{13}
\nonumber\\
&+\cdots.
\end{align}

Thirdly, we illustrate the CPV mechanism with $N\pi$ scatterings.  
Since the transition form factors of $\Lambda_b\to N\pi$ have not been calculated up to now, we will only consider the terms with $\gamma_\mu$ and $\gamma_\mu\gamma_5$.
For example, the transition form factors of $\Lambda_b$ into $J^P=1/2^\pm$ baryons are dominated by the terms of $\bar u_{1/2^+}(f_1^{1/2^+} \gamma_\mu+g_1^{1/2^+}\gamma_\mu\gamma_5)u_{\Lambda_b}$ and $-\bar u_{1/2^-}(f_1^{1/2^-} \gamma_\mu+g_1^{1/2^-}\gamma_\mu\gamma_5)u_{\Lambda_b}$ \cite{Han:2022srw,Huang:2022lfr}, respectively, for the V-A current.  
The minus sign for the $1/2^-$ process is from the negative parity, with the definition of the form factors $f_1^{1/2^-}$ and $g_1^{1/2^-}$ as positive values.
The form factors with higher-spin $N\pi$ states have similar properties. 

Except for neglecting the contributions from other terms of form factors, we also ignore the sub-leading diagrams such as non-factorizable diagrams in which $h^-$ could also re-scatter with $N$ or $\pi$.  
The purpose of this paper is to find a mechanism which can drive the CPV of baryon decays in the low mass region of $p\pi$ and $p\pi\pi$, especially for generating large regional CPV. 
Therefore, we investigate the main mechanism and consider only the dominant contributions, using the above approximations without big problems and leaving the ignored effects in the further discussions in the future. 

Now we can express the full decay amplitude of $\Lambda_b\to (\mathcal{B}M)h^-$ with $\mathcal{B}=p(\Delta^{++})$ and $M=\pi^0(\pi^-)$ as 
{\small
\begin{align}
&\mathcal{A}(\Lambda_b\to (\mathcal{B}M) h^-) 
\nonumber\\
=&
\lambda_u f_h\bar u_{N\pi}  \bigg[
a_1\left( P_{11} f_1^{1/2^+} -S_{11} f_1^{1/2^-} + \cdots \right)m_-
\nonumber\\
&~~~~~~~~~~+a_1\left( P_{11} g_1^{1/2^+} -S_{11} g_1^{1/2^-}+ \cdots \right)m_+\gamma_5
\bigg]
u_{\Lambda_b}
\nonumber\\
+&\lambda_t f_h \bar u_{N\pi}  \bigg[
\left( a_{46+} P_{11} f_1^{1/2^+} - a_{46-}S_{11} f_1^{1/2^-}  + \cdots  \right)m_-
\nonumber\\
&~~~~~~~~+\left(  a_{46-} P_{11} g_1^{1/2^+} - a_{46+}S_{11} g_1^{1/2^-} + \cdots \right)m_+\gamma_5
\bigg]
u_{\Lambda_b}
\end{align}
}
where $\lambda_i=V_{ib}V_{iq}^*$ with $i=u,t$ and $q=d,s$, $f_h$ is the decay constant of $h=\pi$ or $K$, $a_{46\pm}=a_4 \pm R_h a_6$, $R_h=2m_h^2/(m_b(m_u+m_q))$, and $a_{1,4,6}$ are the effective Wilson coefficients in the weak Hamiltonian \cite{Ali:1998eb}, $m_\pm=m_{\Lambda_b}\pm m_{N\pi}$.

It can be seen from the above formula that the interference effects in the tree and penguin contributions are different. 
The above difference comes from the signs of the $a_6$ terms, compared to those of the $a_1$ and $a_4$ terms. 
In the four-fermion operators of $O_{1-4}$, quarks are all left-handed due to the weak interaction $(\bar u b)_{V-A}=\bar u_L\gamma^\mu b_L$. 
On the contrary, the $u$ quark from the $b$ transitions is right-handed with the operators of $O_{5,6}$ after the Feirz transformation, $(\bar u b)_{S-P}=\bar u_R b_L$.
The different chirality of the $u$ quark results in different helicity of the $N\pi$ system in the relativistic case, and hence different interference effects of partial-wave amplitudes. 

There are two sources of difference of strong phases. 
One is from the effective Wilson coefficients of $a_4$ and $a_6$ where the quark loops, the chromo-meganetic operator, and the vertex corrections could contribute the strong phases in the generalized factorization hypothesis \cite{Ali:1998eb}.
It basically contributes to the global CPV.
The second source of difference of strong phases is from the different interfering effects between the partial-wave amplitudes of $N\pi$ scatterings. 
This mainly leads to the regional CPV whose values might be significantly large in some regions of the Dalitz plot where the difference of strong phases could be very large. 

The two sources of CP violation can be clearly illustrated through an analytical example involving two partial-wave amplitudes $S_{11}$ and $P_{11}$, whereas CPV with  higher partial waves are very similar. The partial width with only these terms is given by
\begin{equation}\label{eq:dGammadcostheta}
	\begin{aligned}
d\Gamma&\propto |P_{11}|^{2}(|A|^{2}+\kappa^{2}|B|^{2})+|S_{11}|^{2}(|\tilde A|^{2}+\kappa^{2}|\tilde B|^{2})\\
&+2\mathcal{R}e\left[(A\tilde A^{*}+\kappa^{2}B\tilde B^{*})P_{11}S^{*}_{11}\right]\\
	\end{aligned}
\end{equation}
with 
\begin{equation}
	\begin{aligned}
A&=(\lambda_{u}a_{1}-\lambda_{t}a_{46+})f^{\frac{1}{2}+}_{1}m_{-}\\
B&=(\lambda_{u}a_{1}-\lambda_{t}a_{46-})g^{\frac{1}{2}+}_{1}m_{+}\\
\tilde A&=(-\lambda_{u}a_{1}+\lambda_{t}a_{46-})f^{\frac{1}{2}-}_{1}m_{-}\\
\tilde B&=(-\lambda_{u}a_{1}+\lambda_{t}a_{46+})g^{\frac{1}{2}-}_{1}m_{+}\\
	\end{aligned}
\end{equation}
which can be derived directly from Eq.(4). 
The parameter $\kappa=p_{N\pi}/(E_{N\pi}+m_{N\pi})$, with $p_{N\pi}$, $E_{N\pi}$ and $m_{N\pi}$ as the momentum, energy and mass of the $N\pi$ system in the rest frame of $\Lambda_b^0$. 

In the first line of  Eq. (\ref{eq:dGammadcostheta}), the CP asymmetries are the direct CPV generated by $|A|^2$, $|B|^2$, $|\tilde A|^2$ and $|\tilde B|^2$, as the squared modulus terms of $|P_{11}|^{2}$ and $|S_{11}|^{2}$ do not contribute to interference effects. For example,  
\begin{equation}
	\begin{aligned}
|A|^{2}-|\bar{A}|^{2}&\propto 2 \mathcal{R}e(\lambda_{u}\lambda_{t}a_{1}a_{46+})-2 \mathcal{R}e(\lambda^{*}_{u}\lambda^{*}_{t}a_{1}a_{46+})\\
&\propto \sin(\Delta\phi_{w})\sin(\Delta\delta),
	\end{aligned}
\end{equation}
where the weak phase difference $\Delta\phi_{w}$ stems from the CKM matrix elements $\lambda_{u}\lambda_{t}$ and $\lambda^{*}_{u}\lambda^{*}_{t}$, while the strong phase difference  $\Delta\delta$ is from the effective Wilson coefficients $a_{1}$ and $a_{46+}$. 
CP violation in this case is induced by the interference between the tree and penguin amplitudes in the same partial wave, such as $A$ in the above formula. 

The second line of Eq. (\ref{eq:dGammadcostheta}) introduces a different CPV mechanism where the re-scattering strong phases of $S_{11}$ and $P_{11}$ make the contribution to the strong phase difference of CPV. The associated CP asymmetry, for example, is
\begin{equation}
	\begin{aligned}
&\mathcal{R}e\left[AP_{11}\tilde A^{*}S^{*}_{11}\right]-\mathcal{R}e\left[\bar{A}\bar{P}_{11}\bar{\tilde A}^{*}\bar{S}^{*}_{11}\right]
\\
&\propto \mathcal{R}e\big[(\lambda_u^*\lambda_t-\lambda_u\lambda_t^*)(a_{46+}P_{11})(a_1^*S_{11}^*)\big]
\\
&\hspace{0.4cm}+\mathcal{R}e\big[(\lambda_u\lambda_t^*-\lambda_u^*\lambda_t)(a_{1}P_{11})(a_{46-}^*S_{11}^*)\big].
	\end{aligned}
\end{equation}
The weak phase difference is from $\lambda_u^*\lambda_t$ and $\lambda_u\lambda_t^*$, whereas the strong phase differences originate from $a_{46+}P_{11}$ and $a_1^*S_{11}^*$, or $a_{1}P_{11}$ and $a_{46-}^*S_{11}^*$. The partial-wave amplitudes of $N\pi$ scatterings explicitly contribute to the strong phases. This type of CPV is induced by the interference between the tree and penguin amplitudes in different partial waves. For instance, the tree amplitude of the $1/2^-$ mode, $a_1S_{11}$, interferes with the penguin amplitude of the $1/2^+$ mode, $a_{46+}P_{11}$. Analogous to the interference between the tree amplitude of the $1/2^+$ mode, $a_1P_{11}$, and the penguin amplitude of the $1/2^-$ mode, $a_{46-}S_{11}$.
Besides, since the $N\pi$ scattering amplitudes depend on the colliding energy, the corresponding CPV might vary as the changes of the invariant mass of the $N\pi$ system.

The using of the data of $N\pi$ scatterings is similar to the $\pi\pi\to K\bar K$ rescattering in $B$-meson three-body decays \cite{Bediaga:2013ela,AlvarengaNogueira:2015wpj,Cheng:2016shb,Garrote:2022uub}.
The model-independent dispersive analysis of $\pi\pi\to K\bar K$ data \cite{Pelaez:2020gnd} can lead to a very accurate description of the CP asymmetries of measured by LHCb \cite{Garrote:2022uub}.
However, the CPV mechanism with $N\pi$ scatterings is different from that of $B$-meson decays. 
In $B$-meson decays, the rescattering of $\pi\pi\to K\bar K$ happens between two channels, like $B^\pm\to \pi^\pm \pi^+\pi^-$ and $\pi^\pm K^+K^-$, with the weak phases in these two processes respectively. 
The $N\pi$ scatterings in $\Lambda_b$ decays contribute to one individual process but with different partial waves, which is more similar to the interference of $S$-wave and $P$-wave amplitudes in $B^+\to \pi^+\pi^+\pi^-$. 

\begin{figure}[!t]
    \centering
\includegraphics[scale=0.4]{"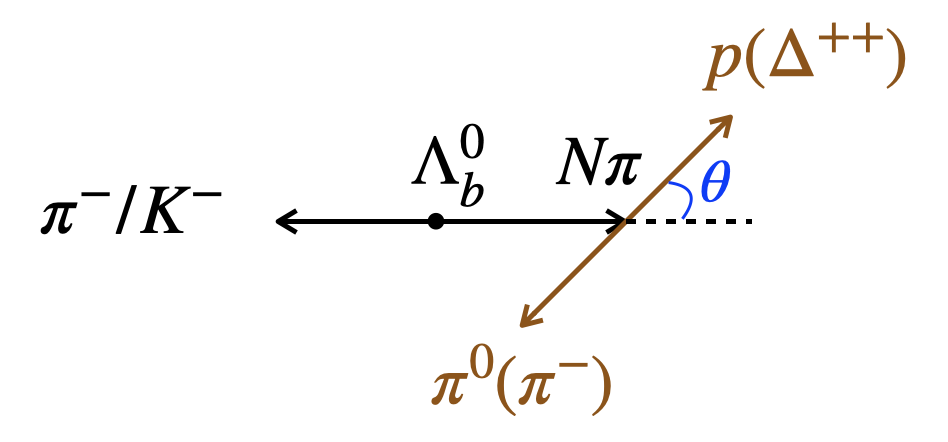"}
        \caption{Angular definition of $\Lambda_b^0\to (p\pi^0)h^-$ and $(\Delta^{++}\pi^-)h^-$ with $h=\pi$ or $K$. }\label{fig:angular}
\end{figure}

In the end, it can be easily studied for the angular analysis on $\Lambda_b$ decays, benefited by the partial-wave amplitudes of $N\pi$ scatterings. 
The angular distributions are illustrated in Fig. \ref{fig:angular} for $\Lambda_b^0\to (p\pi^0)h^-$ and $(\Delta^{++}\pi^-)h^-$. 
The angle $\theta$ is defined between the directions of momentum of the final baryon $p(\Delta^{++})$ in the rest frame of $N\pi$ system and the momentum of $N\pi$ in the rest frame of $\Lambda_b^0$. 
The formula of angular distributions can be easily obtained by the helicity amplitudes. 

The above CPV mechanism from $N\pi$ scatterings can also be applied in charmed baryon decays, such as $\Lambda_c^+\to (p\pi^-)\pi^+$. 
Other scatterings can also generate similar CPV, for instance $N\pi\to \Lambda \bar K$ and $NK\to NK$.
In this work, we will only discuss $\Lambda_b$ decays with $N\pi$ scattering for the introduction of this new CPV mechanism, and leave all the other discussions in the future. 

The proposal of $N\pi$ scatterings can also pave the way for experimental analysis of heavy-flavor baryon decays. 
The amplitude analysis in experiments always suffer large uncertainties and difficulties from the too many baryon resonances.
The using of data of $N\pi$ scatterings might help experimentalists do the partial-wave amplitude analysis. 

%%%%%%%%%%%%%%%%%%%%%%%%%%%%%%%%%%%%%%%%%%%%%%%%
\section{Numerical Discussions}

\begin{table*}[!t]
\centering
\renewcommand\arraystretch{1.8}
\caption{Global and regional CPV of $\Lambda_b$ decays stemming from $N\pi$ re-scatterings. $N\pi\to \Delta\pi$ contributes to the processes of $\Lambda_b^0\to (\Delta^{++}\pi^-)K^-$ and $\Lambda_b^0\to (\Delta^{++}\pi^-)\pi^-$, while $N\pi\to N\pi$ contributes to $\Lambda_b^0\to (p\pi^0)K^-$ and $\Lambda_b^0\to (p\pi^0)\pi^-$. The results are given in three scenarios of the form factors, with S1: $f_{1}=1.1$, $g_{1}=0.9$, S2: $f_{1}=g_1=1.0$, and S3: $f_{1}=0.9$, $g_{1}=1.1$. The predictions are provided in the invariant mass regions of $m_{N\pi}\in [1.2,1.9]$GeV for $N\pi\to \Delta\pi$, and $m_{N\pi}\in [1.1,2.5]$GeV for $N\pi\to N\pi$, which are constrained by the SAID data.}\label{tab:CPV}
\begin{tabular}{ccccc}
\hline\hline
 ~~~~~~decay processes~~~~~~ &~~~~~~ Scenarios ~~~~~~& ~~~~~~~~global CPV ~~~~~~~~ & ~~~~~CPV of $\cos\theta<0$~~~~~~ & ~~~~~~CPV of $\cos\theta>0$~~~~~~ \\
\hline
\multirow{3}{*}{$\Lambda_b^0\to (\Delta^{++}\pi^-)K^-$} & S1 & 5.9\% & 8.0\% & 3.6\%
\\
& S2 & 5.8\% & 6.3\% & 5.3\%
\\
& S3 & 5.6\% & 4.3\% & 7.0\%
\\\hline
\multirow{3}{*}{$\Lambda_b^0\to (\Delta^{++}\pi^-)\pi^-$} & S1 & $-$4.1\% & $-$5.4\% & $-$2.4\% \\
& S2 & $-$3.9\% & $-$3.9\% & $-$3.9\% \\
& S3 & $-$3.6\% & $-$2.3\% & $-$5.3\% \\
\hline
 \multirow{3}{*}{$\Lambda_b^0\to (p\pi^0)K^-$} & S1 & 5.8\% & 8.2\% & 2.7\%
\\
& S2 & 5.8\% & 8.0\% & 3.0\%
\\
& S3 & 5.8\% & 7.8\% & 3.3\%
\\\hline
 \multirow{3}{*}{$\Lambda_b^0\to (p\pi^0)\pi^-$} & S1 & $-$3.9\% & $-$3.9\% &  $-$3.7\%
\\
& S2 & $-$3.9\% & $-$3.8\% &  $-$4.3\%
\\
& S3 & $-$3.8\% & $-$3.6\% &  $-$4.8\%
\\
\hline\hline
\end{tabular}
\end{table*}
\begin{figure*}[!t]
    \centering
    \includegraphics[scale=0.35]{"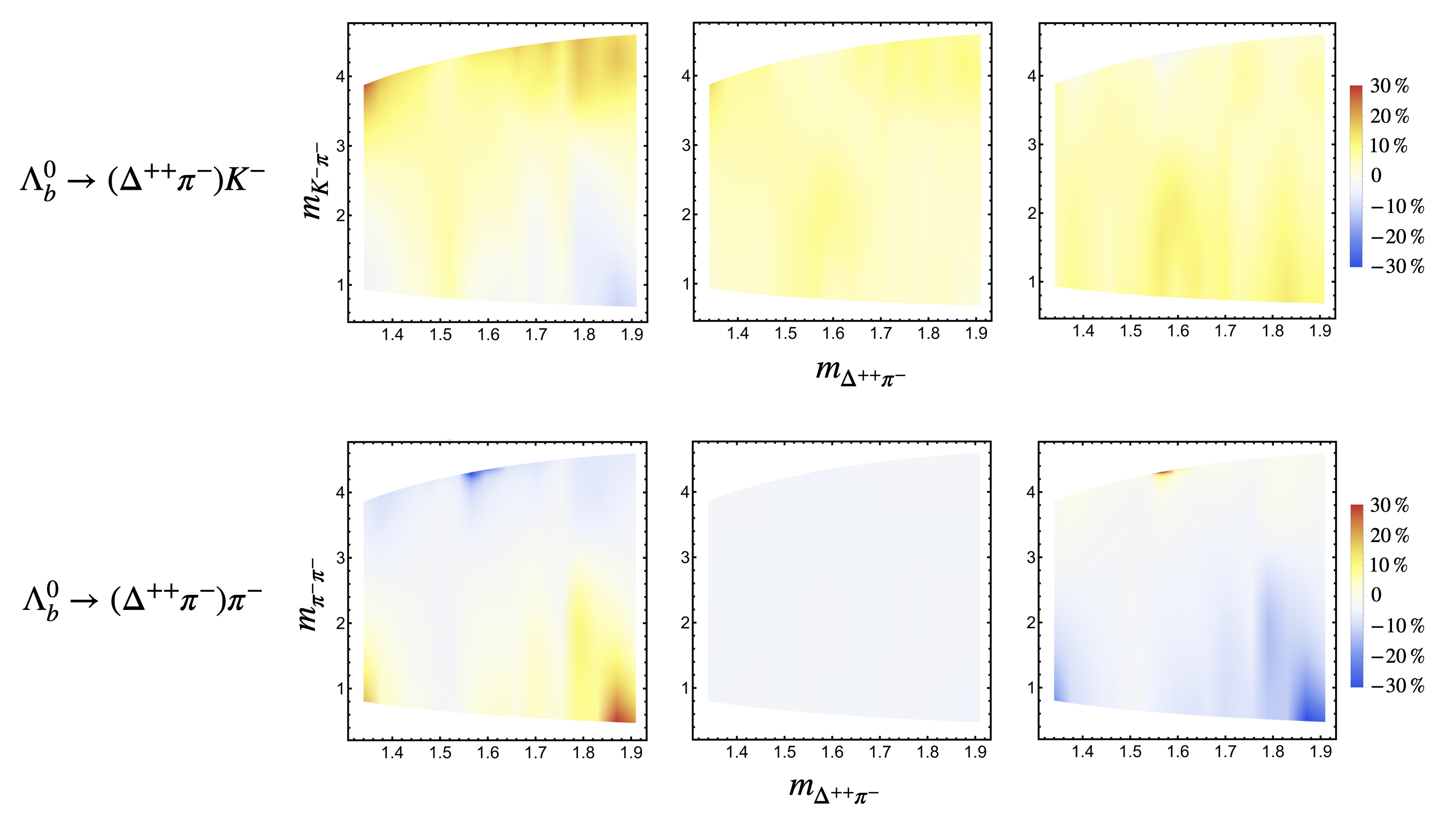"}
    \includegraphics[scale=0.35]{"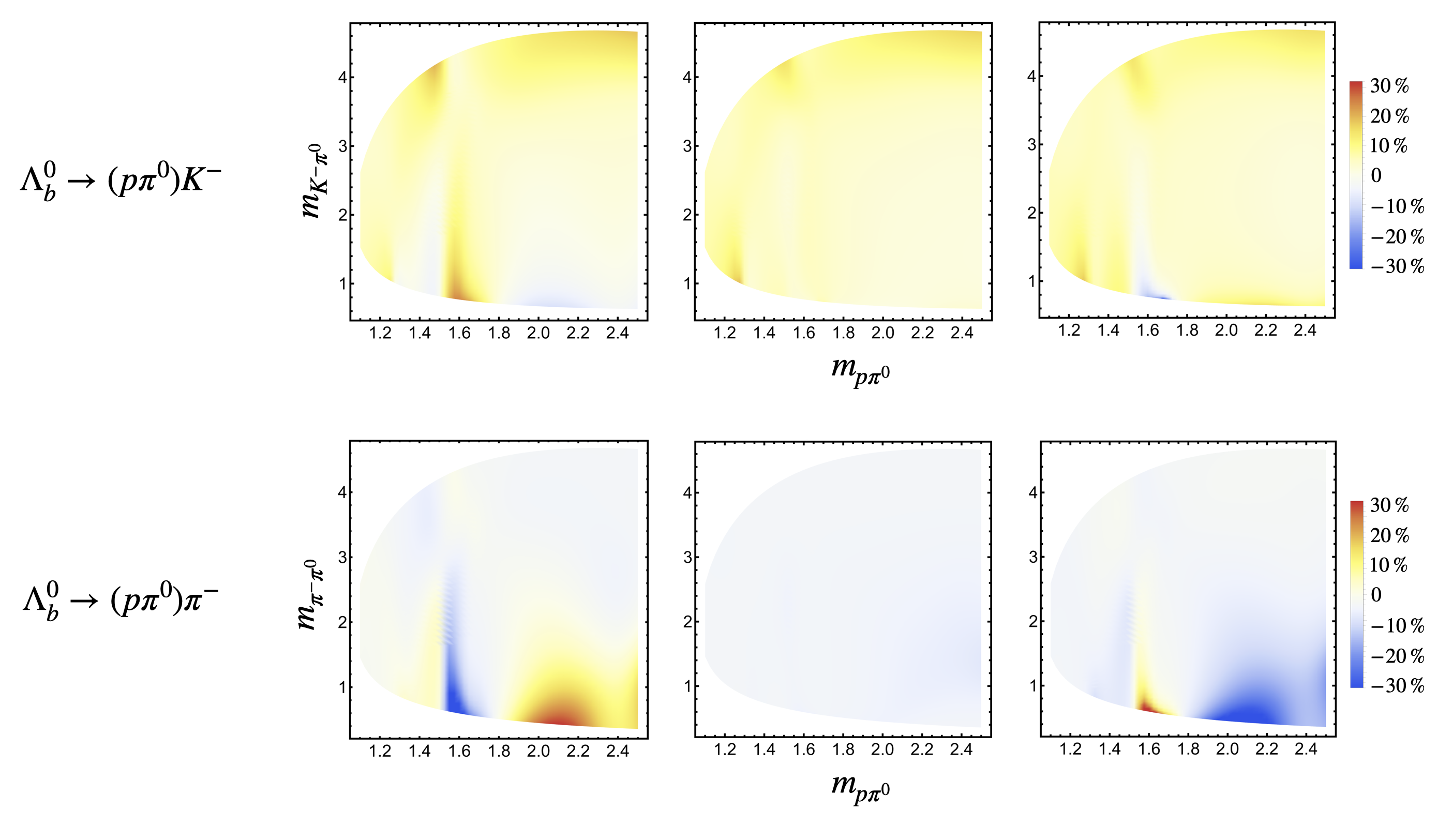"}
    \caption{CPV distributions of the Dalitz plots. They are the processes of $\Lambda^{0}_{b}\to (\Delta^{++}\pi^{-})K^{-}$, $\Lambda^{0}_{b}\to (\Delta^{++}\pi^{-})\pi^{-}$, $\Lambda^{0}_{b}\to (p\pi^{0})K^{-}$ and $\Lambda^{0}_{b}\to (p\pi^{0})\pi^{-}$, from top to bottom. The results for each process are given in three scenarios of the form factors, with S1: $f_{1}=1.1$, $g_{1}=0.9$, S2: $f_{1}=g_1=1.0$, and S3: $f_{1}=0.9$, $g_{1}=1.1$, for the  left, the middle and the right figures, respectively.}\label{fig:Dalitz}
\end{figure*}

Based on the above CPV mechanism and the data of $N\pi$ scatterings, we can predict the global and regional CP asymmetries of $\Lambda_b^0\to (p\pi^0)h^-$ and $(\Delta^{++}\pi^-)h^-$. 
The inputs of effective Wilson coefficients are given in \cite{Ali:1998eb}. 
The inputs of $N\pi$ scattering amplitudes are model-independent and given in \cite{SAID}. 
Since the data of scatterings of $N\pi\to N\pi$ and $\Delta \pi$ are very precise, we neglect the uncertainties of scattering amplitudes. 

\begin{figure*}[!t]
    \centering
    \includegraphics[scale=0.35]{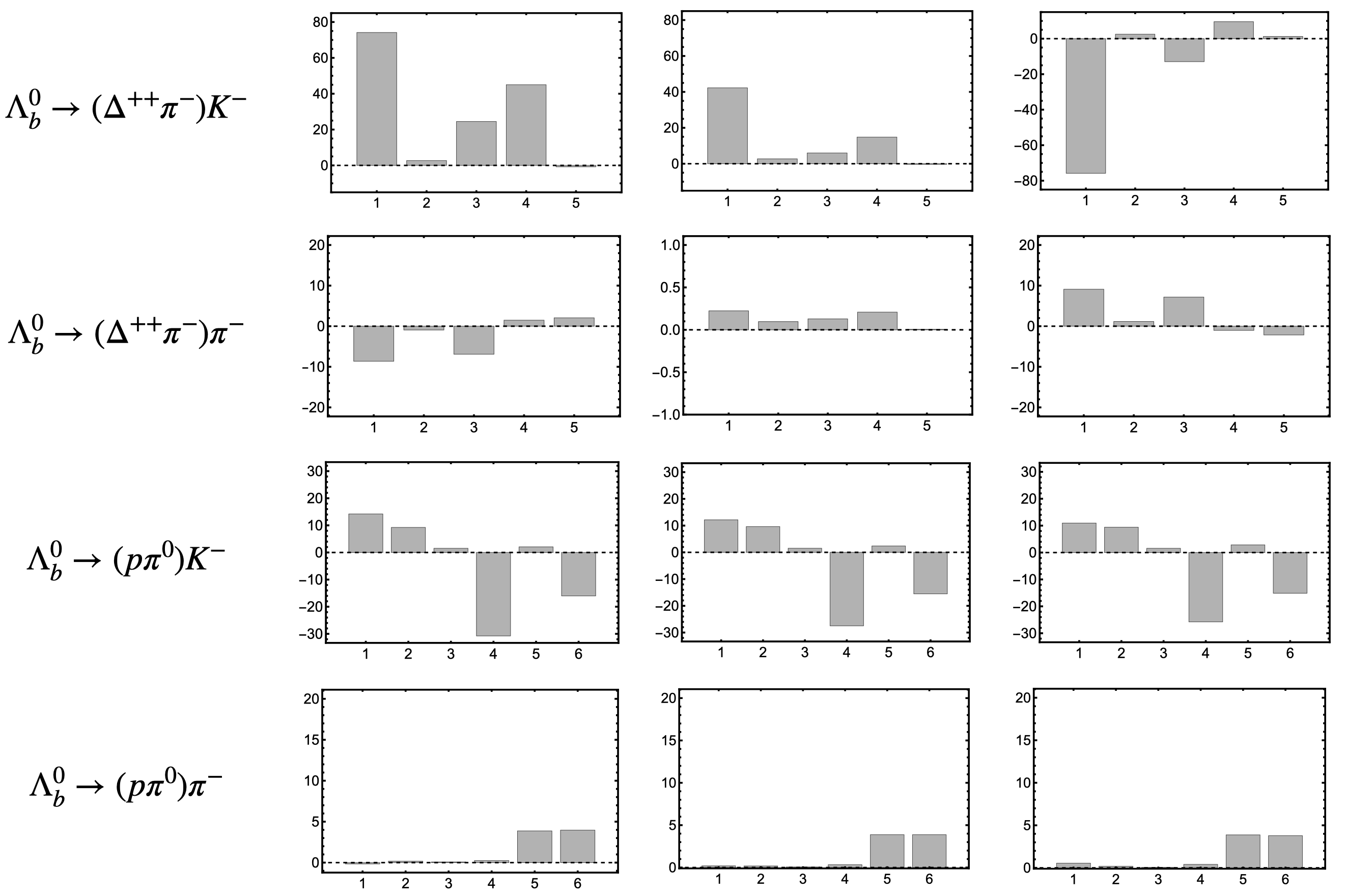}
        \caption{CP asymmetries of the Legendre moments in the unit of percent, with the horizontal axis $n=1,2,\cdots$, of $P_{n}(\cos\theta)$.  They are the processes of $\Lambda^{0}_{b}\to (\Delta^{++}\pi^{-})K^{-}$, $\Lambda^{0}_{b}\to (\Delta^{++}\pi^{-})\pi^{-}$, $\Lambda^{0}_{b}\to (p\pi^{0})K^{-}$ and $\Lambda^{0}_{b}\to (p\pi^{0})\pi^{-}$, from top to bottom. The results for each process are given in three scenarios of the form factors, with S1: $f_{1}=1.1$, $g_{1}=0.9$, S2: $f_{1}=g_1=1.0$, and S3: $f_{1}=0.9$, $g_{1}=1.1$, for the  left, the middle and the right figures, respectively.}\label{fig:Legendre}
\end{figure*}

Little is known about the form factors of $\Lambda_b\to N\pi$ at the current stage. 
But it is similar to the transition form factors of $B\to \pi\pi$ which have been studied by the heavy meson chiral perturbation theory \cite{Lee:1992ih,Fajfer:1998yc,Cheng:2007si,Cheng:2020ipp} and the light-cone QCD sum rules \cite{Cheng:2017smj,Cheng:2017sfk}. 
Therefore the form factors of $\Lambda_b\to N\pi$ can be calculated using the proven methods in principle. 
At the current stage, we will take some simple and typical values of these form factors, in order to reveal the new CPV mechanism. 
In the heavy quark limit, the form factors $f_1=g_1$. 
The values of $f_1$ and $g_1$ might be different by about $10\%$ when considering the corrections of heavy quark expansion \cite{Han:2022srw,Manohar:2000dt}.
Besides, the absolute values of form factors are cancelled in the ratio of CPV definition.
CPV relies only on the relative values of form factors. 
Without loss of generality, we naively take three sets of values, i.e. $f_1=g_1=1.0$ to give results as the central values, with those of $f_1=1.1$, $g_1=0.9$, and $f_1=0.9$, $g_1=1.1$ as the error estimations. 
In the following discussions, we will provide all the results of Table. \ref{tab:CPV} and Figs. \ref{fig:Dalitz} and \ref{fig:Legendre} in three scenarios of the form factors, with S1: $f_{1}=1.1$, $g_{1}=0.9$, S2: $f_{1}=g_1=1.0$, and S3: $f_{1}=0.9$, $g_{1}=1.1$. 
This can be seen what are the results in the heavy quark symmetry, and how large the corrections beyond the heavy quark symmetry could affect. 
The mass regions of $p \pi^0$ and $\Delta^{++}\pi^-$ are taken as what SAID program provides, i.e. $m_{p\pi^0}\in [1.1, 2.5]$GeV and $m_{\Delta^{++}\pi^-}\in [1.2, 1.9]$GeV.
Then we can get the numerical results using the above inputs.

Firstly, the global CPV in the mass regions of the $N\pi$ scattering system are given in Tables. \ref{tab:CPV}.
It can be found that the magnitudes of global CPV of these processes are around $\pm(4-6)\%$. 
Such values of CPV could be measured by LHCb for $\Lambda_b^0\to (p\pi^+\pi^-)K^-$ and $(p\pi^+\pi^-)\pi^-$ whose data samples are large enough with high detection efficiency of all the charged final-state particles.  
From Table. \ref{tab:CPV}, the results of global CPVs do not change significantly with the varying values of form factors $f_1$ and $g_1$. 
The reason is that the global CPV is an overall and averaged effect, and thus is basically generated by the short-distance contributions where the strong phases are from the effective Wilson coefficients. 
The partial-wave amplitudes of the long-distance $N\pi$ scatterings usually affect the local CPVs in the Dalitz plots. 
Therefore, the predictions of the global CPVs can be used for larger regions to be compared with the experimental measurements. 
Even though the invariant mass regions are limited by the SAID data, the global CPV can be extended to larger regions and then compared to the LHCb measurements. 
The LHCb collaboration has measured the CPVs using the data sample of Run I as, $A_{CP}(\Lambda_b^0\to \Delta^{++}\pi^-\pi^-)=(+0.1\pm3.2\pm0.6)\%$ and $A_{CP}(\Lambda_b^0\to \Delta^{++}\pi^-K^-)=(+4.4\pm2.6\pm0.6)\%$ \cite{LHCb:2019jyj}. Our predictions, $A_{CP}(\Lambda_b^0\to (\Delta^{++}\pi^-)\pi^-)\approx-4\%$ and $A_{CP}(\Lambda_b^0\to (\Delta^{++}\pi^-)K^-)\approx 5\%$, are consistent with the LHCb results within the uncertainties. 

The opposite signs between the CPVs of emitted $K^-$ modes and $\pi^-$ modes are basically from the weak phases. The CKM matrix elements of emitted $K^-$ modes are $V_{ub}V_{us}^*$ for the tree contributions, and $V_{tb}V_{ts}^*$ for the penguin contributions. Therefore the CPVs are proportional to $\sin(\gamma)>0$ with $\gamma\sim 65^\circ$ \cite{ParticleDataGroup:2024cfk}. For the emitted $\pi^-$ modes, the CKM matrix elements are  $V_{ub}V_{ud}^*$  and $V_{tb}V_{td}^*$ in the tree and penguin contributions, respectively. The corresponding CPVs are proportional to $\sin(-\alpha)<0$ with $\alpha\sim 84^\circ$ \cite{ParticleDataGroup:2024cfk}.

The regional CPV in the Dalitz plots are even more interesting. 
Every partial-wave amplitude corresponds to different angular distribution. 
Then the interference effects in the different regions in the Dalitz plots would induce different CP asymmetries. 
The numerical results of regional CPV are given in Figure. \ref{fig:Dalitz}. 
It could be found that the local CPV in some regions could reach the order of $10\%$, or be even as large as $30\%$. 
This is large enough for the observation in experiments. 
Besides, the form factors of $f_1$ and $g_1$ could affect the results significantly in some processes, which means the regional CPV are more sensitive to the corrections of heavy quark expansion, compared to the global CPV in Table. \ref{tab:CPV}.

It is also interesting that there exists large regions with CPV of positive or negative values in the Dalitz plot. 
This would be helpful for the experimental measurements. 
For example, it can be divided into two bins with $\cos\theta>0$ and $\cos\theta<0$ in all the four processes. 
The numerical results of CPV in each bin are given in Tables. \ref{tab:CPV}. 

From the Dalitz plots, it could be clearly seen that the regional CPV significantly depend on $\cos\theta$ \footnote{The invariant mass squares, $m_{p\pi^{0}}^2$ or $m_{\Delta^{++}\pi^{-}}^2$, related to the vertical axis in the Dalitz plots, are linearly proportional to $\cos\theta$. Taking $\Lambda_b^0\to (p\pi^0)\pi^-$ as an example, 
\begin{equation}
\begin{aligned}
 m^{2}_{\pi^{0}\pi^{-}}=&\left(\frac{m^{2}_{N\pi}+m^{2}_{\pi^{0}}-m^{2}_{p}}{2m_{N\pi}}+\frac{m^{2}_{\pi^{-}}+m^{2}_{\Lambda_b}-m^{2}_{N\pi}}{2m_{\Lambda_b}}\right)^{2}
 \nonumber\\
 &-\Big(|\vec{p}_{\pi^{0}}|^{2}+|\vec{p}_{\pi^{-}}|^{2}+2|\vec{p}_{\pi^{0}}||\vec{p}_{\pi^{-}}|\cos\theta \Big).
\end{aligned}
\end{equation}
Therefore, the regions of $\cos\theta>0$ and $\cos\theta<0$ are depicted as lower and upper zone in the Dalitz plot, respectively. }.  
This is reasonable that the interference between different partial-wave amplitudes induce the dependence of CPV on $\cos\theta$. 
In order to see the dependence, it is very convenient to express the differential decay width on Legendre series $P_{n}(\cos\theta)$. 
It is known that the angular distributions are actually related to the Legendre polynomials. 
The Legendre distributions after integrating out scattering energy are
\begin{equation}\label{eq:moment}
	\begin{aligned}
\frac{d\Gamma}{d\cos\theta}\propto \sum_{n=0}\mathcal{L}_{n}P_{n}(\cos\theta)
	\end{aligned}
\end{equation}
with $\mathcal{L}_{n}=(1, -0.10, 0.20, -0.05, 0.009, 0.05)$ for $\Lambda^{0}_{b} \to (\Delta^{++}\pi^{-})K^{-}$, $(1,-0.39,0.24,-0.08,-0.09,0.05)$ for $\Lambda^{0}_{b}\to (\Delta^{++}\pi^{-})\pi^{-}$, 
$(1, -0.4, 0.4, -0.5, -0.03, -0.12, $ $-0.005)$ for $\Lambda_b^0\to (p\pi^0)K^-$, and $(1, -1.25, 1.1, -0.7, $ $0.2, -0.1, 0.01)$ for $\Lambda_b^0\to (p\pi^0) \pi^-$,
respectively, in the case of $f_1=1.1$ and $g_1=0.9$.
It can be found that the first few moments make large contributions, with small contributions from higher moments.
This can be understood that the higher partial-wave amplitudes are suppressed, while the lower partial-wave amplitudes make the dominant contributions. 
Then the large regional CPV could also be measured by the Legendre moments. 
It can be easily realized by the experiments using the differential width convoluted with the Legendre polynomials. 
The results of CPV of Legendre moments are given in Figure. \ref{fig:Legendre}. 
The interesting result is that the CPV of the first moment $\cos\theta$ could be as large as $\pm 70\%$ or $40\%$ in $\Lambda_b^0\to (\Delta^{++}\pi^-)K^-$, which can be easily measured and tested in experiments.

In the end, we strongly suggest to measure the CPV of $\Lambda_b^0\to (p\pi^+\pi^-)K^-$ for the first observation of baryon CPV. $\Lambda_b^0\to p\pi^+\pi^-K^-$ has the largest data sample of charmless $b$-baryon decays at LHCb \cite{LHCb:2019jyj}.
It could be measured in three ways. 
Firstly, the global CPV with the value of $6\%$ could be reached with all the data sample.
Secondly, the CPV in the Dalitz plot can be measured with the regional CPV of $(10-20)\%$.  
Thirdly, it is also realizable to measure the CPV of moment of $\cos\theta$ with a value as large as $\pm70\%$ or $40\%$. 
Using the data samples of integrated luminosity of 3fb$^{-1}$ at LHCb Run 1, the total CPV of $\Lambda_b^0\to p\pi^+\pi^-K^-$ is measured to be $(+3.2\pm1.1\pm0.6)\%$, while that of $\Lambda_b^0\to \Delta^{++}\pi^-K^-$ is $(+4.4\pm2.6\pm0.6)\%$ \cite{LHCb:2019jyj}. 
The current total data with 9fb$^{-1}$ collected by LHCb Run 1 and 2 would be larger by around five times due to the larger production rate at Run 2.
So the precision of the global CPV in $\Lambda_b^0\to (p\pi^+\pi^-)K^-$ could be around $1\%$ \cite{YanxiZhang}.
LHCb Run 3 is now collecting even more data during 2024 and 2025, with the total integrated luminosity up to 23fb$^{-1}$. 
Therefore, it can be expected that the CPV of $\Lambda_b^0\to (p\pi^+\pi^-)K^-$ could be observed in the very near future, with the global CPV of around $(5-6)\%$ and the experimental precision lower than $1\%$.

%%%%%%%%%%%%%%%%%%%%%%%%%%
\section{Conclusions}
We propose a new CPV mechanism of baryon decays which stems from $N\pi$ scatterings. 
The weak phases come from the tree and penguin operators, while the strong phases are from the $N\pi$ scatterings. 
The most important advantage is that the fruitful data of $N\pi$ scatterings with precise partial-wave amplitudes could help us do model-independent analysis on the strong phases of $\Lambda_b$ decays. 
This is beneficial for our dynamical predictions on CPV of baryon decays. 
We study the CPV of $\Lambda_b^0\to (p\pi^0)\pi^-$, $(p\pi^0)K^-$, $(p\pi^+\pi^-)\pi^-$ and $(p\pi^+\pi^-)K^-$, via $N\pi\to p\pi^0$ and $N\pi \to \Delta^{++}\pi^-\to p\pi^+\pi^-$, respectively. 
It is found that the global CPV of these processes are at the order of several percents. 
The regional CPV of Dalitz plots could reach as large as $10\%$, or be even larger. 
The CPV of the Legendre moments could be large as well.
Such remarkable CPV have a large possibility to be observed in experiments. 
We strongly suggest to measure the CPV of $\Lambda_b^0\to (p\pi^+\pi^-)K^-$ to achieve the first observation of baryon CPV.

\vspace{0.3cm}
%\section{Acknowledgement}
\begin{acknowledgments}
The authors are very grateful to Yanxi Zhang for the fruitful discussions on the experimental measurements at LHCb, and also to Jinlin Fu, Xiaorui L\"u, Wenbin Qian and Yangheng Zheng for the discussions on the prospects of experimental measurements.
The authors are also grateful to Dan Guo and Igor Strakovsky for the introductions on the data of $N\pi$ scatterings from the SAID program \cite{SAID}, to Hai-Yang Cheng for the discussions on three-body $B$-meson decays, to Shan Cheng and Yu-Ming Wang for the discussions on the $\Lambda_b\to N\pi$ transition form factors, and to Cai-Dian L\"u for the discussions on the CPV from final-state interactions.  
This work is supported in part by Natural Science Foundation of China under
grant No. 12335003, and by the Fundamental Research Funds for the Central Universities under No. lzujbky-2024-oy02 and lzujbky-2023-it12.
\end{acknowledgments}

%%%%%%%%%%%%%%%%%%%%%%%%%%%%%%%
\vspace{0.5cm}
\appendix
\section{Data of $N\pi$ scatterings}\label{sec:SAID}
\begin{figure*}[!t]
    \centering
    \includegraphics[scale=0.5]{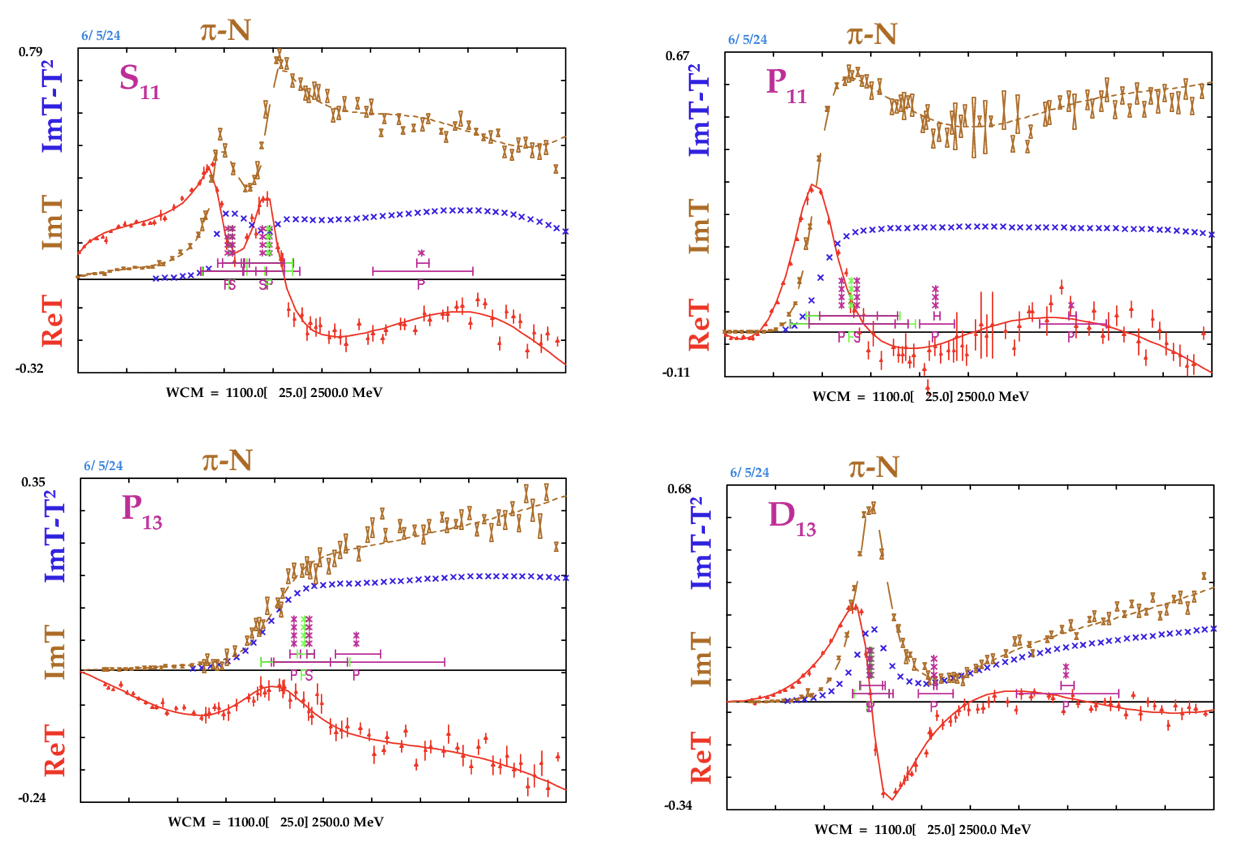}
        \caption{The partial-wave amplitudes of $N\pi\to N\pi$ scatterings, $S_{11}$, $P_{11}$, $P_{13}$ and $D_{13}$. These figures are all taken from the SAID program \cite{SAID}.  }\label{fig:SAID}
\end{figure*}

The partial-wave amplitudes of $N\pi$ scatterings are shown explicitly for the convenience of readers. 
As an example, it is shown for the partial-wave amplitudes of $N\pi\to N\pi$ scatterings, $S_{11}$, $P_{11}$, $P_{13}$ and $D_{13}$.
They are taken from the SAID program \cite{SAID}. See Fig. \ref{fig:SAID}.

%%%%%%%%%%%%%%%%%%%%%%%%%%%%%%%

\end{document}